\documentclass[fleqn,usenatbib]{mnras}

\usepackage[T1]{fontenc}

\DeclareRobustCommand{\VAN}[3]{#2}
\let\VANthebibliography\thebibliography
\def\thebibliography{\DeclareRobustCommand{\VAN}[3]{##3}\VANthebibliography}

\usepackage{graphicx}	
\usepackage{placeins}    
\usepackage{siunitx}		
\usepackage{amsmath}	
\usepackage{amssymb}	
\usepackage{aas_macros}
\usepackage{enumitem} 
\usepackage[shortcuts,acronyms]{glossaries-extra}		

\newabbr{GC}{GC}{globular cluster}
\newabbr[longplural = supernovae, shortplural = SNe]{SN}{SN}{supernova}
\newabbr{IMF}{IMF}{initial mass function}
\newabbr{SF}{SF}{star formation}
\newabbr{MW}{MW}{Milky Way}

\title[Quantifying supernovae from iron spreads]{How many explosions does one need? - Quantifying supernovae in globular clusters from iron abundance spreads}

\author[H. Wirth et al.]{
Henriette Wirth$^{1,2}$\thanks{E-mail: henri-ette\_w@web.de (HW)},
Tereza Jerabkova$^{3}$,
Zhiqiang Yan$^{2,4}$,
Pavel Kroupa$^{2,4}$,
\vspace{5pt}\\
\rm \Large{Jaroslav Haas$^2$
and Ladislav Šubr$^2$}
\\
$^{1}$Elektronische Fahrwerksysteme GmbH, Dr.-Ludwig-Kraus-Str. 6, 85080 Gaimersheim, Germany\\
$^{2}$Charles University, Faculty of Mathematics and Physics, Astronomical Institute, V Hole\v{s}ovi\v{c}kách 2, CZ-18000 Praha 8, Czech Republic\\
$^{3}$European Space Agency (ESA), European Space Research and Technology Centre (ESTEC), Keplerlaan 1, 2201 AZ Noordwijk,\\ The Netherlands\\
$^{4}$Helmholtz Institut für Strahlen und Kernphysik, Universität Bonn, Nussallee 1416, 53115 Bonn, Germany
}

\date{Accepted XXX. Received YYY; in original form ZZZ}

\pubyear{2021}

\begin{document}
\label{firstpage}
\pagerange{\pageref{firstpage}--\pageref{lastpage}}
\maketitle

\begin{abstract}

Many \acp{GC} are known to host multiple populations distinguishable by their light-element content.
Less common are \acp{GC} displaying iron abundance spreads which are seen as evidence for enrichment through core collapse \acp{SN}.
We present a simple analytical method to estimate the number of \acp{SN} required to have occurred in a \ac{GC} from its metallicity and iron abundance spread.
We then use this result to estimate how long \ac{SF} lasted to build the \ac{GC}.
We apply our method to up-to-date measurements and find that, assuming the correctness of these measurements, multiple \acp{SN} (up to $10^5$) are required in most \acp{GC} with iron abundance spreads.
The number of \acp{SN} events which contributed to the enrichment of the \acp{GC} studied here is typically a factor of 10 less than the expected number of \acp{SN} in a canonical \ac{IMF}.
This indicates that gas expulsion from the forming \ac{GC} occurred after the first 10 per cent of \acp{SN} exploded.
We compute that for the \acp{GC} typically \ac{SF} ends after only a few Myr (extending up to $\approx 30 ~\rm Myr$ in a few cases).
We also discuss possible improvements of this method and especially its sensitivity to the error of iron abundance measurements of individual stars of a \ac{GC}.
The method presented here can quickly give an estimate for the number of \acp{SN} required to explain the iron abundance spread in a \ac{GC} without the requirement of any hydrodynamical simulations.
\end{abstract}


\glsresetall



\section{Introduction}

Many \acp{GC} are known to host multiple populations of stars. These can be distinguished by their different abundances of the light elements He, C, N, O, Na, Al.
The origins of the differences in individual \acp{GC} remain unclear \citep[see e.g.][and references therein]{2012A&ARv..20...50G,2019A&ARv..27....8G,2020MNRAS.491..440W}.
One of the most common tools to detecting multiple populations are chromosome maps, that are constructed from two-colour diagrams using well-selected filters and combinations thereof together with spectroscopy allowing to efficiently distinguish populations with different light-element abundances \citep{2015ApJ...808...51M, 2016MmSAI..87..303M, 2017MNRAS.464.3636M}.
While most \acp{GC} are considered to be homogeneous as far as their iron content is concerned \citep{2009A&A...508..695C, 2015ApJ...809..128M, 2021arXiv210307014M}, strong evidence for iron abundance spreads has been found in some of them \citep{2000ApJ...534L..83P,2009Natur.462..483F,2016MNRAS.457...51L,2018ApJ...859...81M}.
In the sample used in this work the \acp{GC} with large iron spreads (likely not to be a consequence of measurement errors) are: NGC 5139, NGC 5272, NGC 5286, NGC 5634, NGC 6205, NGC 6273, NGC 6341, NGC 6656, NGC 6715, Terzan 5 and Terzan 8.
The question then becomes: Where do these iron abundance spreads come from?

If we assume that the \acp{GC} initially formed out of a homogeneous gas cloud, no iron abundance spreads would be expected in the stars that formed prior to the first core collapse \acp{SN}.
The most massive stars in the cluster evolve and produce iron before they finally die in a core collapse \ac{SN} \citep{1995ApJS..101..181W,1998A&A...334..505P}.
Through this \ac{SN} a small portion of the iron from the star is released \citep{1995ApJS..101..181W,1998A&A...334..505P,2017ApJ...848...25M}, which then mixes with any still existing surrounding gas.
If \ac{SF} has not ceased at this point, the now iron enriched gas collapses and forms stars with a slightly higher iron content than the previous one.
This would lead to the observed iron abundance spreads, next to spreads of other elements \citep{2003Natur.422..871U}.
\cite{2021arXiv210602075J} estimated the iron spreads expected for different initial conditions.
However, we could also use the knowledge about the observed iron spreads to estimate the number of \acp{SN} that must have exploded
This is especially relevant, since \acp{SN} are seen as responsible for ending \ac{SF} by blowing out the remaining gas from the \ac{GC} \citep{2000MNRAS.319.1047S,2008MNRAS.384.1231B,2012A&A...546L...5K}.
Therefore, we can use this to better understand when \ac{SF} in \acp{GC} ended.
We implicitly assume here that the \acp{GC} formed their stars in subsequent star formation periods.
However, we are not following the AGB scenario proposed by \cite{1981ApJ...245L..79C}, in which the second generation forms only after the \acp{SN} have expelled all the gas.
In the AGB scenario, the gas the second generation forms out of comes from the ejecta of AGB stars and accreted primordial gas.
While the AGB scenario is likely to contribute to the multiple populations problem in \acp{GC}, the process studied here is more similar to the extended \ac{SF} scenario proposed by \cite{2017ApJ...836...80E}.
Here the star formation continues, while the stars in the dense young \ac{GC} lose gas through stellar evolution and mergers.
Particularly important in this context is that discrete sub-populations of stars with different light-element abundances may arise within the so forming \ac{GC} through mergers of massive stars for which the initial binary fraction is expected to be very high \citep{2020MNRAS.491..440W}.
This pollutes the gas in the \ac{GC} from which enriched stars form.
The \ac{GC} formation process studied here allows for \ac{SF} to continue after the first \acp{SN}.
It is yet to be investigated if this is a viable scenario able to explain all chemical abundances in \acp{GC}.

The purpose of this paper is to present a method to estimate the number of \acp{SN} required to produce the observed iron abundance spreads.
In Sec. \ref{sec_methods} we describe the mathematical methods used in this work.
We apply our method to the data given in \cite{2019ApJS..245....5B} in Sec. \ref{sec_res} and discuss how the model could be improved in Sec. \ref{sec_disc}.
We summarise our findings in Sec. \ref{sec_concl}.

\section{Methods}
\label{sec_methods}

\subsection{Calculating the number of SNe required}
\label{sec_SNreq}

In the scenario studied here the \ac{GC} initially forms from one homogeneous gas cloud i.e. the iron abundance is the same for all stars.
The most massive stars turn into \acp{SN} and pollute the gas that remained in the cluster with iron.
If stars form from this gas, stars with different iron abundances appear in one \ac{GC}.
An observed iron spread implies that \ac{SF} must have been going on after the first \acp{SN} exploded, which is at least for $3 ~\mathrm{Myr}$, but may be longer depending on the upper mass limit of the \ac{GC} stars \citep{2012A&A...537A.146E}.
For our calculations we assume that  the \ac{GC} does not loose any gas or stars before the iron enriched generation has formed and that the gas is well mixed, i.e. the polluted gas becomes homogeneous before new stars form.

To calculate the number of \acp{SN} required to reproduce the iron abundance spreads observed in the \acp{GC} we first compute the total amount of iron that needs to be created.
We assume that the initial metallicity of the \acp{GC} was $([Fe/H] - \sigma_{[Fe/H]})$ and after the gas was enriched by \acp{SN} the metallicity of the remaining gas was $([Fe/H] + \sigma_{[Fe/H]})$, where $[Fe/H]$ is the mean iron abundance of the \ac{GC} and $\sigma_{[Fe/H]}$ the width of the Gaussian fitted over the iron distribution as given in \cite{2019ApJS..245....5B}.
For more information on the data used see Sec. \ref{sec_Bailin}.
We use twice the Gaussian width as our spread, since we assume the metallicity of the first generation to be more towards the lower end of the distribution, which would make the spread larger.
This assumes that the mean metallicity of the \ac{GC} is exactly in the middle of the metallicities of the two generations.
For a \ac{GC} with a pre-\acp{SN}-enriched generation that is larger than the post-\acp{SN}-enriched generation the average metallicity of the \ac{GC} would be lower and since we are using a logarithmic scale, we would underestimate the amount of iron required and vice versa.
This could be improved using detailed histograms showing the metallicities of the stars of the \ac{GC} to determine the metallicities of the individual generations.

Additionally, we neglect any gas loss and assume that the gas is well mixed.
The mass of iron in $M_\odot$ contained in a star of $1 ~M_\odot$ with metallicity $[Fe/H]$ is:
\begin{align}
    m_{\rm iron, star} = p_{\rm iron,\odot} 10^{[Fe/H]},
\end{align}
with $p_{\rm iron,\odot}$ being the mass fraction of iron relative to the total mass of the Sun.
$10^{[Fe/H]}$ produces the mass fraction of iron with respect to hydrogen over that of the Sun.
However, the hydrogen fraction in the \ac{GC} stars are not necessarily the same.
Therefore an additional factor would be needed to account for this discrepancy.
For our estimate, however, we assume that all stars have the same hydrogen content and set this factor to $1$.
Note that the mass fraction of iron with respect to hydrogen, $[Fe/H]$, equals the number fraction usually measured.
$10^{(\log(M_{\rm Fe,star}/M_{\rm H,star}) - log(M_{\rm Fe,\odot}/M_{H,\odot}))}  = (N_{Fe,star}/N_{H,star})/(N_{Fe,\odot}/N_{H,\odot})   = 10^{\rm [Fe/H]}$ since the atomic masses cancel out.

To compute the total mass of iron needed we simply substract the iron in $1~M_\odot$ of the first generation from that of $1~M_\odot$ of the second generation and scale the difference up to the total mass of the gas left over in the \ac{GC}:
\begin{align}
    \label{eq_Miron}
    \begin{split}
        M_{\rm iron} =& ~p_{\rm iron,\odot} \left( 10^{[Fe/H] + \sigma_{[Fe/H]}} - 10^{[Fe/H] - \sigma_{[Fe/H]}} \right) \\
        & \times M_{\rm ecl} \left( \frac{1}{\epsilon} - 1 \right),
    \end{split}
\end{align}
with $M_{\rm iron}$ being the total amount of iron to be created by the \acp{SN} in a \ac{GC} to account for its stellar iron spread, $M_{\rm ecl}$ the initial mass of the stars in the still gas-embedded \ac{GC} and the \ac{SF} efficiency is $\epsilon = \frac{M_{\rm ecl}}{M_{\rm gas}}$, where $M_{\rm gas}$ is the gas mass of the molecular cloud core in which the \ac{GC} forms.
In this work we assume $\epsilon = 0.3$ \citep[see e.g.][]{2003ARA&A..41...57L,2014prpl.conf...27A,2016AJ....151....5M,2018ASSL..424..143B} and $p_{\rm iron,\odot} = 0.0013$ calculated from the measurements given in \cite{2009ARA&A..47..481A}.
This equation can be rewritten an:
\begin{align}
\label{eq_approxMiron}
    M_{\rm iron} = p_{\rm iron, \odot} 10^{[Fe/H]} \times 2 \sinh( \ln(10) \sigma_{\rm [Fe/H]} ) M_{\rm ecl} \left( \frac{1}{\epsilon} - 1 \right).
\end{align}
$\sinh(x) \approx x$ for small values of $x$.
This means that we expect for $M_{\rm iron}$ and as we will see the number of \acp{SN} to scale linearly with $\sigma_{[Fe/H]}$ as long as $\sigma_{[Fe/H]}$ remains small.

Taking the iron yield for a core collapse \ac{SN} to be $0.074 M_\odot$ \citep{2017ApJ...848...25M} and dividing the total amount of iron through this value gives an estimate of the number of \acp{SN} required:
\begin{align}
    \label{eq_NumSN}
    N_{\rm SN} = \frac{M_{\rm iron}}{0.074 M_\odot}.
\end{align}
We used the values for $[Fe/H]$ and $\sigma_{[Fe/H]}$ from \cite{2019ApJS..245....5B} in this work.

Based on the gas expulsion computations by \cite{2017A&A...600A..49B} we assume that the GC looses a negligible fraction of its stars when its residual gas is expelled, i.e. $M_{\rm ini} = M_{\rm ecl}$ to a good approximation.
To calculate $M_{\rm ini}$ for each \ac{GC} in our sample we use Eq. 10 from \cite{2003MNRAS.340..227B}, which describes the dissolution time of the \ac{GC} depending on the initial number of stars, and their Eq. 12, which describes how the \ac{GC} mass changes over time.
These give us:
\begin{align}
     0 = \beta \left[ \frac{\frac{M_{\rm ini}}{\left<m\right>}}{\ln( \gamma \frac{M_{\rm ini}}{\left<m\right>} )} \right]^x \frac{R_{\rm ap}}{\mathrm{kpc}} (1 - e) \frac{1-\frac{M(t)}{0.7 M_{\rm ini}}}{\frac{t}{\mathrm{Myr}}} - 1.
\end{align}
We assume constant parameters $\beta = 1.91$, $x = 0.75$, a Coulomb logarithm, $\gamma = 0.02$, and a mean stellar mass of $\left< m \right> = 0.579 ~M_\odot$, which matches the mean stellar mass for the canonical \ac{IMF} \citep{2001MNRAS.322..231K,2013pss5.book..115K}.
$\beta = 1.91$, $x = 0.75$ are fitting parameters of the model from \cite{2003MNRAS.340..227B}, which depend on the central concentration.
We assumed a rotational velocity of the Galaxy of $220 ~\rm km~s^{-1}$ \citep{1979ApJ...231L.115B,1997PASJ...49..453H}.
The values used here were found for \acp{GC} with a King concentration parameter of $5.0$ on circular orbits.
We took the current mass $M(t)$ and orbit parameters of the \acp{GC} from the \ac{GC} catalog compiled by \cite{hilker_baumgardt_sollima_bellini_2019}.
$R_{\rm ap}$ is the apocentre of the \ac{GC} and $e$ its eccentricity, which we calculated using apo- and pericentre of the \ac{GC}.

The \ac{GC} age $t$ is assumed to be ${12 ~\rm Gyr}$ in accordance with literature values \citep{2010ApJ...708..698D,2019MNRAS.490..491U}.
\cite{2010ApJ...708..698D} looked at 60 \ac{MW} \acp{GC}.
They derived their results by first measuring the magnitude of the main-sequence turnoff and applied isochromes depending on age and metallicity.
This resulted in all but two of their \acp{GC} having ages between 10 and 14 Gyr (See their Table 2).
\cite{2019MNRAS.490..491U} fit photometric measurements with synthetic models for 83 \ac{MW} \acp{GC}.
They derive \ac{GC} ages between $5.5$ and $14.5 ~\rm Gyr$.
All the parameters are, therefore, estimated reasonably well and we solve this equation implicitly for $M_{\rm ini}$ for each \ac{GC} using the Newton-Raphson method.

\subsection{Computing when star formation ends}

If we know how many \acp{SN} have to explode for the next generation to form, we can use this to estimate the time until \ac{SF} ends.
We assume, that gas mixing and \ac{SF} happens instantaneously.
This means that \ac{SF} ends after the $N_{\rm SN}$th \ac{SN} took place and before the $(N_{\rm SN} + 1)$th happenes.
We use the canonical \ac{IMF} \citep{2001MNRAS.322..231K,2013pss5.book..115K} for stellar masses between $0.08 ~M_\odot$ and $m_{\rm max} = 120 ~M_\odot$ to compute the number of \acp{SN} expected to happen.
The minimum mass for a \ac{SN} progenitor is assumed to be $m_{\rm SN} = 8 ~M_\odot$.
From this we can calculate that about ${0.6 ~\%}$ of the cluster stars are more massive than ${8 ~M_\odot}$ which means that they will turn into \acp{SN}.
The possibility of failed \acp{SN} as discussed in \cite{2015ApJ...801...90P,2016ApJ...821...38S,2020arXiv200715658B,2021arXiv210403318N} is neglected here (for more discussion see Sec. \ref{sec_expSN}).
Therefore, the expected number of \acp{SN}, $N_{\rm SN}^{\rm exp}$ is:
\begin{align}
    N_{\rm SN}^{\rm exp} = 0.006 \frac{M_{\rm ini}}{\left< m \right>} = 0.0109 \frac{M_{\rm ini}}{M_\odot}.
\end{align}
We use the integral of the upper end of the \ac{IMF}:
\begin{align}
N_{\rm SN}^{\rm exp} = \int\limits_{m_{SN}}^{m_{max}} dm ~ k \xi(m),
\end{align}
where $k$ is obtained by normalising the \ac{IMF} to $M_{\rm ini}$.
$\xi(m) = m^{-\alpha_i}$ is the canonical \ac{IMF} with $\alpha_1 = 1.3$ for $0.08\le m/M_\odot < 0.5$ and $\alpha_2=2.3$ for $0.5\le m/M_\odot \le 120$.
With this we can compute the mass of the $N_{\rm SN}$th \ac{SN}, $m_{\rm last}$, to explode:
\begin{align}
    \frac{N_{\rm SN}}{N_{\rm SN}^{\rm exp}} = \frac{m_{\rm max}^{-1.3} - m_{\rm last}^{-1.3}}{m_{\rm max}^{-1.3} - m_{\rm SN}^{-1.3}}.
\end{align}
It is assumed here that the iron output of \acp{SN} is the same for all stars (i.e. no mass dependency).
We use the correlation between $m_{\rm last}$ and its lifespan shown in fig. 3 of \cite{2019A&A...629A..93Y} for [Fe/H] = -1.67 to compute the time until \ac{SF} ends.
Their fig. 3 shows how the lifespan of stars decreases with increasing stellar mass.
Most of our \acp{GC} have a low metallicity and the lifespans of stars above $8 ~M_\odot$ do not vary significantly at low metallicity.

\subsection{The dataset used}
\label{sec_Bailin}

The catalogue by \cite{2019ApJS..245....5B} can be seen as an extension of the catalogue from \cite{2012AJ....144...76W}.
It is the largest catalogue of iron spreads in \acp{GC} to date.
They combine several studies of red giant branch measurements in \acp{GC}.
Combining different data sources like this produces an artificially inflated $\sigma_{[Fe/H]}$.
\cite{2019ApJS..245....5B} used a Bayesian Markov Chain Monte Carlo method to correct for this.

They also argue that the error estimates in some of the studies they use are overestimated and correct for this.
A discussion of the dataset and the impact of those errors is given in Sec. \ref{sec_expSN} and App. \ref{sec_systErr}.

\section{Results}
\label{sec_res}

\begin{table*}
    \caption{The current mass, $M(t)$, initial mass, $M_{\rm ini}$, metallicity $[Fe/H]$, iron abundance spread $\sigma_{[Fe/H]}$, number of SNe per unit mass, $n_{\rm SN} = \frac{N_{\rm SN}}{M_{\rm ini}}$, number of SNe, $N_{\rm SN}$, the expected number of SNe to occur in the GC, $N_{\rm SN}^{\rm exp}$, the time SF ends, $t_{\rm end}^{\rm SF}$, and the number of SNe after correcting for the measurement error, $N_{\rm SN}^{\rm subt}$, calculated for our sample of GCs.}
    \begin{tabular}{lrrrrrrrrr}\hline
Name & {$M(t)$} & {$M_{\rm ini} [10^5 M_\odot]$} & {$[Fe/H]$} & {$\sigma_{[Fe/H]}$} & {$n_{\rm SN} [M_\odot^{-1}]$} & {$N_{\rm SN}$} & {$N_{\rm SN}^{\rm exp} $} & {$t_{end}^{SF} [{\rm Myr}]$} & {$N_{\rm SN}^{\rm subt}$} \\
\hline
47 Tuc&     8.07&    12.85&   -0.747&    0.033&{$1.18\times 10^{-3}$}&{$1.52\times 10^{3}$}&{$1.33\times 10^{4}$}&      5.6&{$-6.27\times 10^{2}$}\\
NGC 288&     0.98&     2.88&   -1.226&    0.037&{$4.39\times 10^{-4}$}&{$1.27\times 10^{2}$}&{$2.99\times 10^{3}$}&      4.1&{$-3.32\times 10^{1}$}\\
NGC 362&     3.36&     8.69&   -1.213&    0.074&{$9.09\times 10^{-4}$}&{$7.89\times 10^{2}$}&{$9.00\times 10^{3}$}&      5.0&{$2.93\times 10^{2}$}\\
NGC 1851&     2.81&     8.48&   -1.157&    0.046&{$6.41\times 10^{-4}$}&{$5.43\times 10^{2}$}&{$8.79\times 10^{3}$}&      4.5&{$-7.74\times 10^{0}$}\\
NGC 1904&     1.56&     6.46&   -1.550&    0.027&{$1.52\times 10^{-4}$}&{$9.81\times 10^{1}$}&{$6.69\times 10^{3}$}&      3.6&{$-7.18\times 10^{1}$}\\
NGC 2419&    14.30&    20.79&   -2.095&    0.032&{$5.14\times 10^{-5}$}&{$1.07\times 10^{2}$}&{$2.15\times 10^{4}$}&      3.5&{$-4.94\times 10^{1}$}\\
NGC 2808&     8.69&    17.45&   -1.120&    0.035&{$5.30\times 10^{-4}$}&{$9.25\times 10^{2}$}&{$1.81\times 10^{4}$}&      4.2&{$-3.09\times 10^{2}$}\\
NGC 3201&     1.41&     2.41&   -1.496&    0.044&{$2.81\times 10^{-4}$}&{$6.76\times 10^{1}$}&{$2.49\times 10^{3}$}&      3.8&{$-4.15\times 10^{0}$}\\
NGC 4590&     1.32&     2.06&   -2.255&    0.053&{$5.89\times 10^{-5}$}&{$1.21\times 10^{1}$}&{$2.13\times 10^{3}$}&      3.5&{$1.43\times 10^{0}$}\\
NGC 4833&     1.80&     6.76&   -2.070&    0.013&{$2.21\times 10^{-5}$}&{$1.49\times 10^{1}$}&{$7.01\times 10^{3}$}&      3.4&{$-3.89\times 10^{1}$}\\
NGC 5024&     4.17&     6.47&   -1.995&    0.071&{$1.44\times 10^{-4}$}&{$9.32\times 10^{1}$}&{$6.71\times 10^{3}$}&      3.6&{$3.2\times 10^{1}$}\\
NGC 5053&     0.73&     1.35&   -2.450&    0.041&{$2.91\times 10^{-5}$}&{$3.91\times 10^{0}$}&{$1.39\times 10^{3}$}&      3.4&{$-5.56\times 10^{-1}$}\\
NGC 5139&    33.40&    53.55&   -1.647&    0.271&{$1.3\times 10^{-3}$}&{$6.96\times 10^{3}$}&{$5.55\times 10^{4}$}&      5.9&{$5.83\times 10^{3}$}\\
NGC 5272&     3.60&     5.95&   -1.391&    0.097&{$7.93\times 10^{-4}$}&{$4.72\times 10^{2}$}&{$6.17\times 10^{3}$}&      4.8&{$2.46\times 10^{2}$}\\
NGC 5286&     3.59&     8.51&   -1.727&    0.103&{$3.89\times 10^{-4}$}&{$3.31\times 10^{2}$}&{$8.82\times 10^{3}$}&      4.0&{$1.82\times 10^{2}$}\\
NGC 5466&     0.55&     1.09&   -1.865&    0.075&{$2.05\times 10^{-4}$}&{$2.23\times 10^{1}$}&{$1.13\times 10^{3}$}&      3.6&{$8.44\times 10^{0}$}\\
NGC 5634&     2.20&     3.98&   -1.869&    0.081&{$2.2\times 10^{-4}$}&{$8.74\times 10^{1}$}&{$4.12\times 10^{3}$}&      3.7&{$3.72\times 10^{1}$}\\
NGC 5694&     3.67&     6.13&   -2.017&    0.046&{$8.84\times 10^{-5}$}&{$5.42\times 10^{1}$}&{$6.35\times 10^{3}$}&      3.5&{$-8.95\times 10^{-1}$}\\
NGC 5824&     8.49&    12.58&   -2.174&    0.058&{$7.78\times 10^{-5}$}&{$9.78\times 10^{1}$}&{$1.30\times 10^{4}$}&      3.5&{$1.90\times 10^{1}$}\\
NGC 5904&     3.68&     6.55&   -1.259&    0.041&{$4.51\times 10^{-4}$}&{$2.96\times 10^{2}$}&{$6.79\times 10^{3}$}&      4.1&{$-4.10\times 10^{1}$}\\
NGC 5986&     3.31&    11.19&   -1.527&    0.061&{$3.63\times 10^{-4}$}&{$4.06\times 10^{2}$}&{$1.16\times 10^{4}$}&      3.9&{$9.56\times 10^{1}$}\\
NGC 6093&     2.64&    17.90&   -1.789&    0.014&{$4.54\times 10^{-5}$}&{$8.13\times 10^{1}$}&{$1.86\times 10^{4}$}&      3.5&{$-1.91\times 10^{2}$}\\
NGC 6121&     0.93&     7.53&   -1.166&    0.050&{$6.82\times 10^{-4}$}&{$5.14\times 10^{2}$}&{$7.81\times 10^{3}$}&      4.5&{$3.45\times 10^{1}$}\\
NGC 6139&     3.48&     8.66&   -1.593&    0.033&{$1.68\times 10^{-4}$}&{$1.46\times 10^{2}$}&{$8.98\times 10^{3}$}&      3.6&{$-6.08\times 10^{1}$}\\
NGC 6171&     0.81&     4.50&   -0.949&    0.047&{$1.06\times 10^{-3}$}&{$4.76\times 10^{2}$}&{$4.67\times 10^{3}$}&      5.4&{$3.78\times 10^{0}$}\\
NGC 6205&     4.69&     9.58&   -1.443&    0.101&{$7.33\times 10^{-4}$}&{$7.02\times 10^{2}$}&{$9.92\times 10^{3}$}&      4.6&{$3.8\times 10^{2}$}\\
NGC 6218&     0.87&     2.84&   -1.315&    0.029&{$2.80\times 10^{-4}$}&{$7.96\times 10^{1}$}&{$2.94\times 10^{3}$}&      3.8&{$-4.87\times 10^{1}$}\\
NGC 6229&     2.88&     5.78&   -1.129&    0.044&{$6.54\times 10^{-4}$}&{$3.78\times 10^{2}$}&{$5.99\times 10^{3}$}&      4.5&{$-2.28\times 10^{1}$}\\
NGC 6254&     1.89&     4.90&   -1.559&    0.049&{$2.71\times 10^{-4}$}&{$1.32\times 10^{2}$}&{$5.08\times 10^{3}$}&      3.8&{$6.24\times 10^{0}$}\\
NGC 6266&     6.90&    17.29&   -1.075&    0.041&{$6.89\times 10^{-4}$}&{$1.19\times 10^{3}$}&{$1.79\times 10^{4}$}&      4.6&{$-1.65\times 10^{2}$}\\
NGC 6273&     6.57&    14.18&   -1.612&    0.161&{$8.03\times 10^{-4}$}&{$1.14\times 10^{3}$}&{$1.47\times 10^{4}$}&      4.8&{$8.15\times 10^{2}$}\\
NGC 6341&     3.12&     8.43&   -2.239&    0.083&{$9.61\times 10^{-5}$}&{$8.1\times 10^{1}$}&{$8.73\times 10^{3}$}&      3.5&{$3.55\times 10^{1}$}\\
NGC 6362&     1.16&     3.21&   -1.092&    0.017&{$2.75\times 10^{-4}$}&{$8.82\times 10^{1}$}&{$3.33\times 10^{3}$}&      3.8&{$-1.54\times 10^{2}$}\\
NGC 6366&     0.43&     2.25&   -0.555&    0.071&{$3.97\times 10^{-3}$}&{$8.92\times 10^{2}$}&{$2.33\times 10^{3}$}&     13.3&{$3.08\times 10^{2}$}\\
NGC 6388&    11.30&    21.88&   -0.428&    0.054&{$4.03\times 10^{-3}$}&{$8.82\times 10^{3}$}&{$2.27\times 10^{4}$}&     13.5&{$1.22\times 10^{3}$}\\
NGC 6397&     0.90&     2.62&   -1.994&    0.028&{$5.67\times 10^{-5}$}&{$1.49\times 10^{1}$}&{$2.72\times 10^{3}$}&      3.5&{$-9.98\times 10^{0}$}\\
NGC 6402&     7.39&    19.12&   -1.130&    0.053&{$7.86\times 10^{-4}$}&{$1.50\times 10^{3}$}&{$1.98\times 10^{4}$}&      4.8&{$1.81\times 10^{2}$}\\
NGC 6441&    12.50&    24.24&   -0.334&    0.079&{$7.35\times 10^{-3}$}&{$1.78\times 10^{4}$}&{$2.51\times 10^{4}$}&     25.6&{$7.35\times 10^{3}$}\\
NGC 6535&     0.13&     3.65&   -1.963&    0.035&{$7.61\times 10^{-5}$}&{$2.78\times 10^{1}$}&{$3.78\times 10^{3}$}&      3.5&{$-9.35\times 10^{0}$}\\
NGC 6553&     4.45&    11.23&   -0.151&    0.047&{$6.64\times 10^{-3}$}&{$7.45\times 10^{3}$}&{$1.16\times 10^{4}$}&     22.6&{$7.45\times 10^{1}$}\\
NGC 6569&     2.28&     6.09&   -0.867&    0.055&{$1.49\times 10^{-3}$}&{$9.11\times 10^{2}$}&{$6.31\times 10^{3}$}&      6.4&{$1.39\times 10^{2}$}\\
NGC 6626&     2.84&    12.60&   -1.287&    0.075&{$7.77\times 10^{-4}$}&{$9.79\times 10^{2}$}&{$1.31\times 10^{4}$}&      4.7&{$3.72\times 10^{2}$}\\
NGC 6656&     4.09&     7.39&   -1.803&    0.132&{$4.21\times 10^{-4}$}&{$3.11\times 10^{2}$}&{$7.66\times 10^{3}$}&      4.0&{$2.02\times 10^{2}$}\\
NGC 6681&     1.20&     6.26&   -1.633&    0.028&{$1.30\times 10^{-4}$}&{$8.14\times 10^{1}$}&{$6.48\times 10^{3}$}&      3.5&{$-5.46\times 10^{1}$}\\
NGC 6715&    16.20&    23.68&   -1.559&    0.183&{$1.04\times 10^{-3}$}&{$2.46\times 10^{3}$}&{$2.45\times 10^{4}$}&      5.3&{$1.85\times 10^{3}$}\\
NGC 6752&     2.32&     4.81&   -1.583&    0.034&{$1.77\times 10^{-4}$}&{$8.54\times 10^{1}$}&{$4.99\times 10^{3}$}&      3.6&{$-3.20\times 10^{1}$}\\
NGC 6809&     1.88&     5.19&   -1.934&    0.045&{$1.05\times 10^{-4}$}&{$5.44\times 10^{1}$}&{$5.38\times 10^{3}$}&      3.5&{$-2.12\times 10^{0}$}\\
NGC 6838&     0.63&     1.64&   -0.736&    0.039&{$1.43\times 10^{-3}$}&{$2.35\times 10^{2}$}&{$1.70\times 10^{3}$}&      6.2&{$-4.60\times 10^{1}$}\\
NGC 6864&     4.10&     8.05&   -1.164&    0.059&{$8.1\times 10^{-4}$}&{$6.51\times 10^{2}$}&{$8.34\times 10^{3}$}&      4.8&{$1.37\times 10^{2}$}\\
NGC 7078&     4.99&     8.50&   -2.287&    0.053&{$5.48\times 10^{-5}$}&{$4.65\times 10^{1}$}&{$8.80\times 10^{3}$}&      3.5&{$5.47\times 10^{0}$}\\
NGC 7089&     5.20&    15.58&   -1.399&    0.021&{$1.67\times 10^{-4}$}&{$2.61\times 10^{2}$}&{$1.61\times 10^{4}$}&      3.6&{$-3.2\times 10^{2}$}\\
NGC 7099&     1.39&     4.30&   -2.356&    0.037&{$3.26\times 10^{-5}$}&{$1.4\times 10^{1}$}&{$4.45\times 10^{3}$}&      3.5&{$-3.72\times 10^{0}$}\\
Terzan 1&     2.72&    30.29&   -1.263&    0.037&{$4.03\times 10^{-4}$}&{$1.22\times 10^{3}$}&{$3.14\times 10^{4}$}&      4.0&{$-3.20\times 10^{2}$}\\
Terzan 5&     7.60&    17.56&   -0.092&    0.295&{$5.14\times 10^{-2}$}&{$9.02\times 10^{4}$}&{$1.82\times 10^{4}$}& {-} &{$7.70\times 10^{4}$}\\
Terzan 8&     0.58&    25.47&   -2.255&    0.098&{$1.1\times 10^{-4}$}&{$2.79\times 10^{2}$}&{$2.64\times 10^{4}$}&      3.5&{$1.47\times 10^{2}$}\\
\hline
\end{tabular}

    \label{tab_GCProps}
\end{table*}

\begin{figure}
    \includegraphics{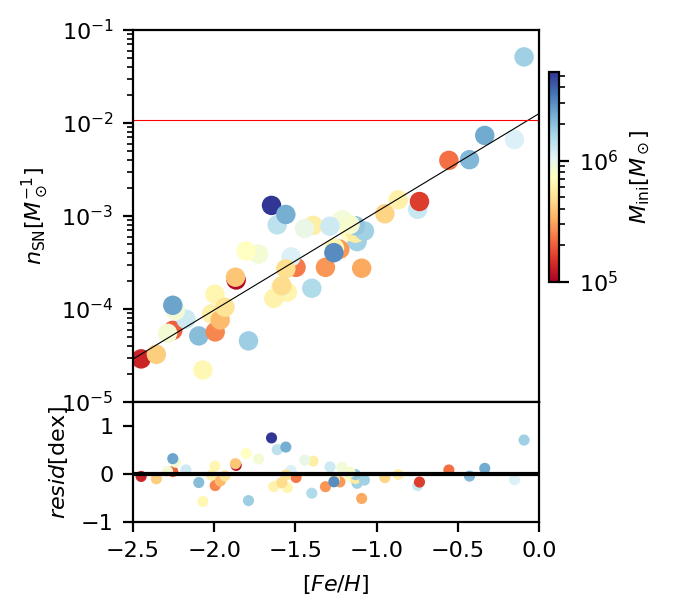}
    \caption{A visualization of the data in Table \ref{tab_GCProps}. The number of SNe per mass is plotted over the metallicity, colourcoded for the initial mass for all 55 GCs.
    The red line shows the maximum number of \ac{SN} per mass possible for the canonical \ac{IMF}.}
    \label{fig_FeH-SNDens}
\end{figure}

\begin{figure}
    \includegraphics{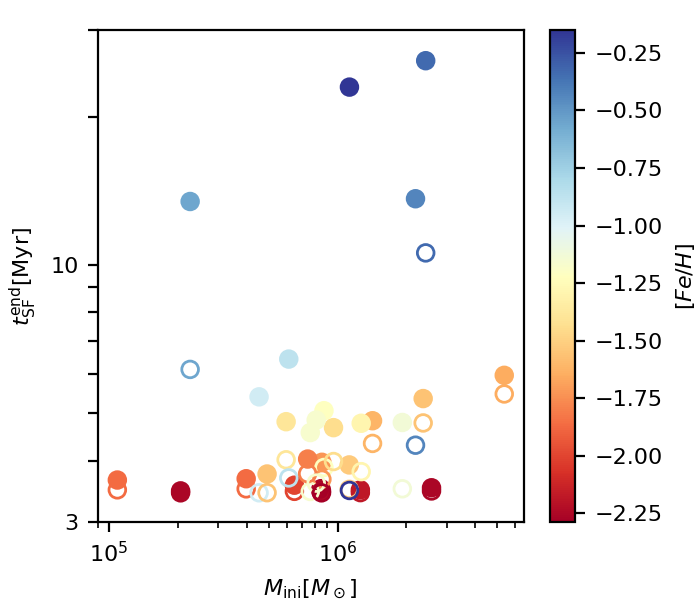}
    \caption{A visualization of the data in Table \ref{tab_GCProps}. The time required for the GCs until SF ends over their initial mass, colourcoded for the metallicity. The filled dots are the values from Table \ref{tab_GCProps}, the empty dots show the time when SF ends for the corrected numbers of SNe, $N_{\rm SN}^{\rm subt}$.}
    \label{fig_Tsf}
\end{figure}

In Table \ref{tab_GCProps} we see the current and initial masses, metallicities $[Fe/H]$, iron abundance spreads $\sigma_{[Fe/H]}$, required number of \acp{SN} per unit mass, $n_{\rm SN}$, required numbers of \acp{SN}, $N_{\rm SN}$, calculated for the iron abundance spreads given in \cite{2019ApJS..245....5B}.
The next columns show the number of \acp{SN} expected to occur in the \ac{GC} given its initial mass, the estimated time at which \ac{SF} ends and the number of \acp{SN} required corrected for the measurement error of metallicity measurements of individual stars.
For the \acp{GC} in our sample, the number of \acp{SN} required to produce the observed iron abundance spreads is between five and a few tens of thousand with \ac{SN} densities between $2.2 \times 10^{-5}$ and $7.4 \times 10^{-3} ~M_\odot^{-1}$.
The exeption is Terzan 5 with a number of \acp{SN} per mass of $5.1 \times 10^{-2} ~M_\odot^{-1}$.
This contradicts previous claims that only a few \acp{SN} can occur before the remaining gas is blown out of the \ac{GC} and \ac{SF} stops \citep{2000MNRAS.319.1047S}.

Fig. \ref{fig_FeH-SNDens} shows $n_{\rm SN}$ from Table \ref{tab_GCProps} over the metallicities of the \acp{GC}.
A positive dependency of $n_{\rm SN}$ on $[Fe/H]$ implied by our analytical model exists for the data in our sample. It is, however, somewhat weaker if we correct for the individual measurement errors for the metallicities of individual stars (see Fig. \ref{fig_FeH-SNDens-error} and Sec. \ref{sec_systErr}).
Contrary to theoretical predictions \citep{2018ApJ...863...99B}, we do not find any correlation between $n_{\rm SN}$ and $M_{\rm ini}$.

Fig. \ref{fig_Tsf} shows $t_{\rm end}^{\rm SF}$ over $M_{\rm ini}$.
As we can see, the majority of \acp{GC} takes around ${4 ~\rm Myr}$ until \ac{SF} ends and $T_{\rm SF}$ is similar for most of them.
According to the data we used from \cite{2019A&A...629A..93Y} a $120 M_\odot$ star, which is the upper limit we used for the \ac{IMF}, explodes after about $3.5 ~\rm Myr$.
Therefore, in most \acp{GC} \ac{SF} stops soon after the first \acp{SN} explode.
The implied dependency from Fig. \ref{fig_FeH-SNDens} between $n_{\rm SN}$ and $[Fe/H]$ leads to a similar dependency between $t_{\rm SF}^{\rm end}$ and $[Fe/H]$.
After correcting for the effect of measurement errors (see Sec. \ref{sec_systErr}) \ac{SF} stops for all \acp{GC} except for NGC 6441 after a similar time.

\section{Discussion}
\label{sec_disc}

\subsection{Comparison with the expected number of SNe}
\label{sec_expSN}

As we can see in Table \ref{tab_GCProps} the expected number of \acp{SN}, $N_{\rm SN}^{\rm exp}$, computed for the canonical \ac{IMF}, is larger (often by one or two orders of magnitude) than the number of \acp{SN}, $N_{\rm SN}$, as computed here, with the exception of Terzan 5.
This indicates that \ac{SF} ends long before most of the \acp{SN} explode.
At the same time, however, the effects neglected in our simplified analytical model may also play a role.
These include a loss of \ac{SN} ejecta due to their high velocities during \ac{GC} formation \citep{2015MNRAS.454.4197R} and the fact that a portion of the most massive stars is not expected to turn into \acp{SN} but become failed \acp{SN} instead \citep{2015ApJ...801...90P,2016ApJ...821...38S,2020arXiv200715658B,2021arXiv210403318N}.
The fraction of those failed \acp{SN} is still unclear, however, the upper limit is estimated to be at around $50 ~\%$ \citep{2010PhRvD..81h3001L,2011ApJ...730...70O,2014MNRAS.445L..99H,2015ApJ...801...90P,2016ApJ...821...38S} and is, therefore, not sufficient to explain the difference between required and expected numbers of \acp{SN} in most \acp{GC}.
Another effect that would reduce the number of \acp{SN} happening in a \ac{GC} is the ejection of O stars due to stellar-dynamical encounters.
However, \cite{2015ApJ...805...92O} found that only about $10 \%$ of O stars would be lost this way for \acp{GC}.
One of the most important unknowns, when it comes to determining the number of \acp{SN} expected, is the exact shape of the \ac{IMF}, since it determines the fraction of stars massive enough to turn into \acp{SN}.
While we used the canonical \ac{IMF} \citep{2001MNRAS.322..231K,2013pss5.book..115K} in this work, several authors found that the shape of the \ac{IMF} varies depending on the \ac{GC} parameters \citep{2006A&A...458..135P,2012MNRAS.422.2246M,2017MNRAS.471.2242B,2017A&A...608A..53J,2018ApJ...857..132K,2018Sci...359...69S,2021arXiv210304997C}.
Additionally, the upper limit of the IMF can vary depending on the \ac{SF} rate \cite[e.g.][]{2013pss5.book..115K,2017MNRAS.471.2242B,2017ApJ...834...94S,2017A&A...607A.126Y}.

The larger $N_{\rm SN}$ for Terzan 5 might be a result of incomplete mixing (see Sec. \ref{sec_desc_limit}) which can cause us to highly overestimate $N_{\rm SN}$ and $n_{\rm SN}$.
It has also been suggested that Terzan 5, similar to $\omega$ Cen, is not a usual \ac{GC}, but the remnant of a larger structure \citep{2014ApJ...795...22M}.

The processes described here do not exclude each other, and it is likely that a multitude of the processes described here worked together to form the \acp{GC} we observe today.

The time when \ac{SF} ends varies between 3.5 and 33.5 Myr with the majorities of cases clustering around 4 Myr.
Since this result relies on the expected number of \acp{SN} it is also likely to be dependent on the upper end of the \ac{IMF}.
Variations in the iron output of \acp{SN} with different progenitor masses also play a role.
If more massive stars (above $m_{\rm last}$) would for example turn out to release twice as much iron as less massive ones only half of the stars would have to explode before the \ac{SF} ends.
Studies investigating the dependence of \ac{SN} yields on mass and metallicity are yet to be performed.

In this paper we assumed that gas mixing and \ac{SF} happen instantaneously after the $N_{\rm SN}$th \ac{SN} explodes.
However, those processes are expected to take some time.
Hydrodynamical simulations would be required to investigate the time required for mixing and the fractions of iron in the newly forming stars due to possible incomplete mixing.
These simulations, are however, difficult and expensive to do.
They would have to take different conditions like the density profile and the presence of dust into account and might even need to include radiative transfer.
This makes the more detailed simulations more challenging and emphasises the usefulness of the method presented here.

\subsection{The limitations of our method}
\label{sec_desc_limit}

The main advantage of our method is that it is purely analytical.
Due to the complexity of the problem hydrodynamical simulations would have to be very detailed, containing a number of effects like \ac{SN} winds from different \acp{SN} reducing or enhancing each other.
Therefore, they would be computationally very expensive to perform.
However, it is important to pay close attention to the assumptions used in our model.

As discussed in Sec. \ref{sec_SNreq} $n_{\rm SN}$ scales approximately linearly with $\sigma_{\rm [Fe/H]}$.
This means that if we used only half the spread for our estimate we would also only get half the $n_{\rm SN}$.
Therefore, an over- or underestimation of $\sigma_{\rm [Fe/H]}$ will influence $n_{\rm SN}$ linearly, too.
For a more detailed discussion on the effect of these errors see App. \ref{sec_systErr}.

As mentioned before, we are using the mean value of $[Fe/H]$ in our model and assume that it is exactly in the middle of the metallicities of both \ac{SN}-enriched and not enriched populations.
This means that if the second generation is comprised of more stars than the first one, we overestimate the number of \acp{SN} required and vice versa.
It is important to note that this depends on the fraction of stars from each population in the sample of stars measured and not in the entire \ac{GC}.
The selection of \ac{GC} stars, therefore, has a strong influence on the result.
\cite{2017MNRAS.464.3636M} found, based on chromosome maps, that the number fraction of pre-\ac{SN}-population stars declined with \ac{GC} mass.
In that case we would expect larger $n_{SN}$ for the more massive \acp{GC}.
This is not the case with our current data.

Similarly to populations being under- or overrepresented, entire populations might be excluded from the sample.
For example \cite{2019ApJS..245....5B} took the data for NGC 7089 from \cite{2019A&A...622A.191M} which according to the author (private communication) did not include stars from the anomalous giant branch discovered by \cite{2014MNRAS.441.3396Y}.
This can lead to an underestimate of the iron spreads.

Next to the systematic error for $[Fe/H]$, which was neglected in the \cite{2019ApJS..245....5B} data, there are several assumptions in our model that could influence our results:
We assume a constant iron output from all our \acp{SN}, but in reality, more massive ejecta are expected at larger stellar masses and metallicities \citep{2019MNRAS.484.2587M}.
However, the chemical composition and mass fraction of \ac{SN} yields is not fully understood yet \citep{2020MNRAS.495.3751C}.
Additionally, the high-mass end of the \ac{IMF} was found to vary with \ac{GC} density and metallicity \citep{2012MNRAS.422.2246M}.
These variations of the \ac{IMF} would affect the composition of \ac{SN} producing stars and therefore the average iron yield of the \acp{SN}.

Furthermore, we assume well mixed gas, i.e., the first generation of stars forms out of a homogeneous gas cloud, then the \acp{SN} polute the gas.
The gas mixes so that the next generation of stars forms out of a homogeneous gas cloud as well.
We also assume that no gas is lost from the \ac{GC}.
It is however not clear how the gas mixes and there is evidence for gas being lost from \acp{GC} \citep{2015ApJ...814L..14C,2016MNRAS.461.4088D,2020MNRAS.493.1306C}.
If the gas is not well mixed pockets of larger metallicity remain in the gas, which would lead to stars with higher iron content and therefore larger iron abundance spreads in the \ac{GC} as a whole.
For example if the \acp{SN} ejecta mix only with half of the gas left over from the first \ac{SF} event, we would only need half of the \acp{SN} to explain the observed iron spreads.
The same would be true, if half of the gas was ejected from the \ac{GC} before enrichment through \acp{SN}.
Due to mass segregation the heavier stars which evolve into \acp{SN} are concentrated in the centre of the \ac{GC}.
Therefore, the gas ejected by them would largely be retained, while blowing out the gas further away from the \ac{GC}'s centre.
This was further investigated by \cite{2021arXiv210602075J}, who found that the amount of gas retained depends on the initial gas mass and the core radius of the \ac{GC}.

We also have to consider the error introduced by $M_{\rm ini}$ since $N_{\rm SN}$ depends linearly on it.
It does not affect $n_{\rm SN}$ since the $M_{\rm ini}$ in $N_{\rm SN}$ cancels with the $M_{\rm ini}$ in the denominator.
In our calculations we assumed the same mean stellar mass, $\left< m \right>$, for all \acp{GC}.
\cite{2012MNRAS.422.2246M} found, however, that the \ac{IMF} varies with the metallicity of the \ac{GC}.
The $M_{\rm ini}$ we calculate is roughly proportional to $\left< m \right>$ and therefore, $N_{\rm SN}$ is roughly proportional to $\left< m \right>$, too.

\subsection{Individual clusters}

In the following sections we discuss individual \acp{GC} for which the number of \acp{SN} required is so large that even if assuming a large error (see App. \ref{sec_systErr}) at least $10^3$ \acp{SN} are required to produce the observed iron spread.
We will discuss if iron spreads for these \acp{GC} have been observed in other studies as well and whether or not spreads in the light-element content of \ac{GC} stars has been observed as well.
Note that spreads in light elements can also be caused from pollution of the \ac{GC} gas before the first \acp{SN} for example through stellar mergers \citep{2020MNRAS.491..440W}.

\subsubsection{NGC 5139}

The most famous \ac{GC} on our list is NGC 5139 or $\omega$ Cen.
The iron spread in this \ac{GC} has been confirmed via chromosom maps by \cite{2019MNRAS.487.3815M}.
They identify three distinctive populations with different mean iron spreads, but also internal spreads.
They find that the iron-rich population is also rich in sodium and nitrogen compared to the iron-poor one, but poor in oxygen.
Since NGC 5139 is an unusual \ac{GC} \citep[very large mass, retrograde orbit, e.g.][]{1999AJ....117.1792D, 2018ApJ...853...86B} and it has been suggested that it might have developed from Galactic threshing \citep{2003MNRAS.346L..11B}.
According to our calculations, star formation in NGC 5139 ends after 6 Myr with the lowest mass star whose \ac{SN} contributes to the iron spread being 34 $M_\odot$.
As mentioned in Sec. \ref{sec_desc_limit}, composition of \ac{SN} ejecta depending on mass and metallicity of the progenitor star is not fully understood.
Therefore, we cannot draw any conclusion what consequences the \ac{SN} ejecta might have on the light element content.

\subsubsection{NGC 6388}

In addition to the measurements featured in \cite{2019ApJS..245....5B}, iron spreads in NGC 6388 are also visible from CaT measurements \citep{2020A&A...635A.114H}.
This method relies on the correlation between the equivalent width of Ca lines and $[Fe/H]$ as found by \cite{1988AJ.....96...92A}.
\cite{2020A&A...635A.114H} find a difference of 0.22 dex between different populations of this \ac{GC} although with large measurement errors.
\cite{2018A&A...614A.109C} claim  for this \ac{GC} to have no iron spread.
Variations in both oxygen and sodium have been detected for this \ac{GC}.
\cite{2007A&A...464..967C} and \cite{2018A&A...614A.109C} also investigated the \ac{GC} for other elements and found Na-O and Mg-Al anticorrelations.
They find an excess of $\alpha$-process elements compared to the surrounding field stars.
However, if the iron measurements are correct we find that \ac{SF} lasts very long in this \ac{GC} (13.6 Myr), therefore we would expect large abundance spread for the lighter elements as well.
A consequence of this long lasting \ac{SF} is that the mass of the last star whose \ac{SN} ejecta contribute is also lower ($16 ~M_\odot$).

\subsubsection{NGC 6441}

Similar to NGC 6388, NGC 6441 was investigated by \cite{2020A&A...635A.114H} using CaT measurements, however, they could not confirm an iron spread due to large measurement uncertainties.
\cite{2007A&A...464..953G}, whose data was used to create our dataset in \cite{2019ApJS..245....5B}, also state that the measured variation in iron does not exceed measurement errors.
Therefore it is likely, that the apparent iron spread in NGC 6441 is only an artefact and more precise observations are needed to confirm whether or not an iron spread in NGC 6441 exists.
If our data is correct, however, \ac{SF} in this \ac{GC} lasts even longer, than in the previous \acp{GC} (26 Myr) which leads to a minimum mass for stars contributing to the iron spread of only $10 ~M_\odot$ which is close to the lower limit of stars turning into \acp{SN}.

\subsubsection{NGC 6715}

\cite{2019MNRAS.487.3815M} used chromosome maps to identify three populations in NGC 6715 and found iron spreads both within and between the different populations.
They also identified a spread in sodium, oxygen and magnesium. According to \cite{2019MNRAS.487.3815M} oxygen and sodium are anticorrelated.
\ac{SF} lasted only for 5.3 Myr in this \ac{GC} and, therefore, the minimum mass for stars contributing to the iron spread is very large ($39 ~M_\odot$).

\subsubsection{Terzan 5}

As described in Sec. \ref{sec_expSN} Terzan 5 is not an ordinary \acp{GC} and was suggested to be the remnant of a larger structure \citep{2014ApJ...795...22M}.
The lack of an Al-O anticorrelation, as is often observed in \acp{GC}, further underlines the unusual status of Terzan 5 \citep{2011ApJ...726L..20O}.
This might be related to the number of $N_{\rm SN}$ being is above the number of $N_{\rm SN}^{exp}$.

\section{Conclusions}
\label{sec_concl}

We presented a new method to estimate the number of \acp{SN} required to reproduce the iron abundance spreads observed in \acp{GC}.
Since it is an analytical method we do not need to run any computationally expensive hydrodynamical simulations.
As observational data improves, this method can be used to yield more precise estimates and adjusted to develop a clearer understanding of what the metallicities of different stellar populations are.

Based on the values given in \cite{2019ApJS..245....5B} we find that multiple \acp{SN} (up to $10^5$) are required to obtain the iron abundance spreads in most \acp{GC}.
We also demonstrate how to estimate the time \ac{SF} ends based on these values.
In all cases (except for Terzan 5) the number of \acp{SN} needed is significantly smaller than the number expected for a canonical stellar population and \ac{SF} ends after a few Myr.

To obtain more reliable results more detailed models including iron yields depending on stellar mass and metallicity are required.
Future work should also include hydrodynamical simulations to investigate the degree of mixing the gas in the \acp{GC} experiences before the second population of stars forms after the first \acp{SN}.
The dependency on the \acp{GC}' \acp{IMF} of different \ac{GC} parameters should also be taken into account.
Additionally, better measurements of present day iron abundance spreads in \acp{GC} need to be done.

\section*{Acknowledgements}

We thank Franti\v{s}ek Dinnbier for our many useful discussions. The authors acknowledge support from the Grant Agency of the Czech Republic under grant number 20-21855S.

\section*{Data availability}

The data used here has been cited and is available in published form.



\bibliographystyle{mnras}
\bibliography{IronSpreadGC} 

\appendix
\FloatBarrier

\section{The sensitivity of our model towards measurement errors}
\label{sec_systErr}

\begin{figure}
    \includegraphics{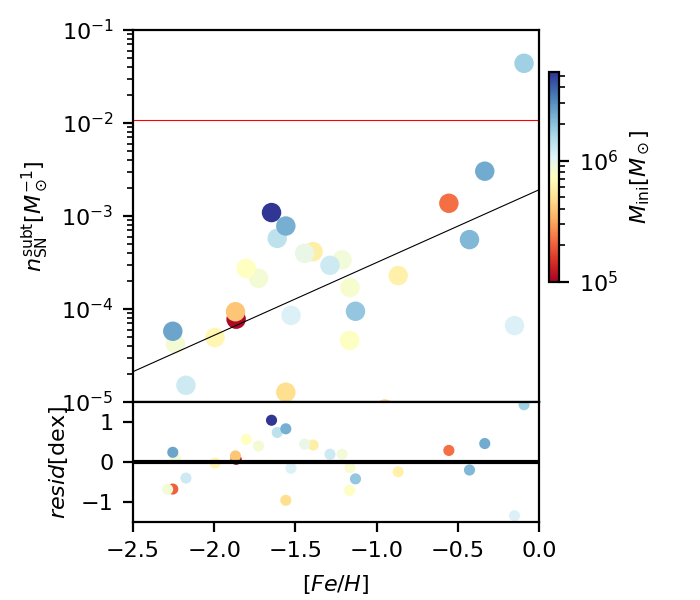}
    \caption{The same as Fig. \ref{fig_FeH-SNDens} but with the error described in Sec. \ref{sec_systErr} substracted.
    Since approximately a third of our GCs end up with negative values for $N_{\rm SN}^{\rm subt}$ only 26 GCs are visible here.
    Three GCs are missing from the top panel because they have positive values below $10^{-5} M_{\odot}^{-1}$.
    Since the fit is done on an exponential scale and the residuals are in dex, only positive values were used here.}
    \label{fig_FeH-SNDens-error}
\end{figure}

Our model is based on the Gaussian fit over the metallicity measurements of individual stars in the \acp{GC}.
As such it is expected to be sensitive to errors of the individual metallicity measurements of those stars since they could stretch the Gaussian, making the iron spread look larger than it actually is.
While \cite{2019ApJS..245....5B} argues that the random error in many of the previous studies have been overestimated (see his sec. 3.2), and therefore applied an independent random error estimation for each individual star using their Fe I lines, however, an underestimation of the random error of his method cannot be excluded based on the observational data available in the error analysis in Bailin's sec. 3.2.
Indeed, it is unavoidable that some random error cannot be accounted for using the Fe I lines taken at the same time at the same spot on a CCD for a single star.
Therefore, we investigate the effect of a larger random error on our result.
We generated a population of synthetic \acp{GC} without any internal iron spread.
Their metallicities ranged from $[Fe/H] = -2.5$ and $-0.1$.
We created a sample of 150 stars for each \ac{GC}, which is consistent with the numbers of stars measured in the \acp{GC} with GIRAFFE \citep{2009A&A...505..117C}.
Since there is no significant dependency between the metallicity values and their error in \cite{2009A&A...505..117C} we assigned their measurement errors to our stars randomly.
We use the error values from \cite{2009A&A...505..117C} because these measurements were also used by \cite{2019ApJS..245....5B}, though \cite{2019ApJS..245....5B} argues that their error estimates are too large.
\cite{2009A&A...505..117C} computed those errors from the errors of the atmospheric parameters of the stars (effective temperatures and microturbulent velocity) and the equivalent widths.
They multiplied these individual errors with the sensitivities of the abundance to variations of the corresponding parameter and then summed them by quadrature.
For more information see \cite{2009A&A...505..117C}.
The error values they computed this way follow a normal distribution. They varied between 0.003 dex and 0.287 dex with a mean of $0.130 ~\rm dex$.

The metallicity values of our synthetic stars were then shifted randomly within their error range, mimicking the effect of the error.
We used Eqs. \ref{eq_Miron} and \ref{eq_NumSN} to compute the apparent $n_{\rm SN}$ for the synthetic \acp{GC}.
We then fitted $n_{\rm SN}$ over $[Fe/H]$ with the following function:
\begin{align}
    n_{\rm SN}^{\rm synt} = a \times e^{b \times [Fe/H]} + c
\end{align}
with $a$, $b$ and $c$ being free parameters and obtain $a = ( 1.3 \pm 0.1 ) \times 10^{-3} ~M_\odot^{-1}$, $b = 2.3 \pm 0.1$ and $c = (0 \pm 2) \times 10^{-5} ~M_\odot^{-1}$.
This shows that the error of our model increases sharply with $[Fe/H]$.

Assuming that the error values for all \acp{GC} are similar to those from \cite{2009A&A...505..117C}, we subtract this from $n_{\rm SN}$ and obtain the results in Table \ref{tab_GCProps} with $n_{\rm SN}^{\rm subt} = n_{\rm SN} - n_{\rm SN}^{\rm synt}$ and $N_{\rm SN}^{\rm subt} = n_{\rm SN}^{\rm subt} \times M_{\rm ini}$.
As we can see $N_{\rm SN}^{\rm subt}$ is very close to zero or negative for a third of the \acp{GC}.
However, for the majority of \acp{GC} a significant number of \acp{SN} is required to explain the measured iron spread even after correcting for the effect of the error of the metallicity measurements of individual stars.

Fig. \ref{fig_FeH-SNDens-error} visualizes the results.
As indicated by its definition $N_{\rm SN}^{\rm subt}$ is smaller than $N_{\rm SN}$.
The implied dependency is also less visible than before, however, it does not disappear.
As discussed in Sec. \ref{sec_res} this also leads to smaller values of $t_{\rm SF}^{\rm end}$.
The larger residuals show that the implied dependency between $n_{\rm SN}$ and $[Fe/H]$ is weaker after the correction.
The same is true for the Pearson coefficient.
For the uncorrected values the Pearson coefficient is 0.91 and for the corrected values only 0.59.
This shows that at least in part the error in the individual measurements of the stars is responsible for the implied dependency.

\bsp	
\label{lastpage}
\end{document}